\documentclass[11pt,a4paper]{article}
\usepackage[T1]{fontenc}

\usepackage[a4paper,top=2cm,bottom=2cm,left=3cm,right=3cm,marginparwidth=1.75cm]{geometry}

\usepackage{amsmath}
\usepackage{graphicx}
\usepackage{hyperref}
\usepackage{siunitx}
\usepackage{parskip}

\usepackage[affil-it]{authblk}

\makeatletter
\renewcommand{\section}{\@startsection {section}{1}{\z@}%
              {24pt}{12pt} {\large\scshape\bfseries}}

\renewcommand{\subsection}{\@startsection {subsection}{2}{\z@}%
             {12pt}{12pt}  {\itshape\bfseries}}

\usepackage{apacite}
\usepackage{natbib}

\usepackage{tikz}

\usetikzlibrary{patterns,decorations.pathreplacing,shapes,arrows.meta,positioning,calc,fit,backgrounds}

\tikzset{
  mainbox/.style={
    rectangle, draw=black, thick,
    minimum width=3.2cm, minimum height=0.75cm,
    text centered, font=\small
  },
  groupbox/.style={
    rectangle, draw=gray, thick,
    inner sep=8pt
  },
  nestedbox/.style={
    rectangle, draw=black, thick,
    minimum width=1.5cm, minimum height=0.65cm,
    text centered, font=\small
  },
  arr/.style={
    -Stealth, thick
  },
  lightarr/.style={
    -Stealth, thin, gray
  },
  label/.style={
    font=\footnotesize\itshape, text=gray
  }
}

\usepackage{enumitem}

\newcommand{\keepcomment}{0} 
\usepackage[normalem]{ulem}
\usepackage{xcolor}
\ifnum\keepcomment=1
    \usepackage[colorinlistoftodos,textsize=scriptsize]{todonotes} 
    \newcommand{\stkout}[1]{\ifmmode\text{\sout{\ensuremath{#1}}}\else\sout{#1}\fi}
    
\else
    
    \usepackage[disable]{todonotes} 
\fi

\newcommand{\todoi}[1]{\todo[inline]{#1}}
\usepackage{tabularx}
\usepackage{subcaption}
\usepackage{amsmath}

\DeclareMathAlphabet{\pazocal}{OMS}{zplm}{m}{n}
\usepackage{amssymb}   
\usepackage{bm}        
\usepackage{algorithm}
\usepackage{algorithmic}
\usepackage{bbm} 
\usepackage{hyperref}

\usepackage{multirow}

\usepackage{layout}

\title{\bfseries \normalsize Accounting for intra-household joint travel in agent-based transport simulations}

\author[1,2]{Lucas Javaudin*}
\author[1]{Andrea Araldo}
\author[2]{Nicolas Coulombel}

\affil[1]{SAMOVAR, Télécom SudParis, Institut Polytechnique de Paris, 91120 Palaiseau, France}
\affil[2]{LVMT, ENPC, Institut Polytechnique de Paris, Univ Gustave Eiffel, 77420 Champs-sur-Marne, France}

\date{\vspace{-5ex}}

\begin{document}
\maketitle

\section*{Short summary}\small

Intra-household joint home-based tours --- trips in which household members depart together, engage in shared activities, and return together --- represent a significant share of daily travel, yet are systematically ignored in transport simulations.
Conflating joint and solo tours within a single mode choice framework introduces bias in preference parameter estimates.
This paper proposes a three-step methodology to integrate joint tours in agent-based transport models: a Random Forest classifier to identify joint tours, a Multinomial Logit model estimating mode choice specific to joint tours, and a Penalized Logistic Regression for driver/passenger assignment.
Applied to the Paris region using household travel survey data, the methodology successfully replicates observed joint tour shares and mode distributions in a synthetic population.
The proposed framework enables more reliable evaluation of policies whose impacts differ between joint and solo travel, such as HOV lanes or family transit fare discounts.

\textbf{Keywords}: agent-based modelling; family carpooling; intra-household interactions; mode choice; synthetic population.

\section{Introduction}

In the Paris region, car trips account for 101 million vehicle-kilometres per day, of which 79 million are driven and 22 million are travelled as a passenger \citep{EGT_2018-2020}.
A notable share of this --- 24 million vehicle-kilometres, comprising 13 million as passenger and 11 million as driver --- involves travelling with at least one other household member, accounting for a total of 2.4 million daily trips.
This form of intra-household ridesharing, variously referred to as family carpooling, family ridesharing, or fampool, thus constitutes a significant component of mobility in the region.

Among the 11 million fampool kilometres travelled as driver, 7.3 million (\SI{67}{\%}) are performed as part of fully-joint home-based tours, defined as sequences of trips that originate and terminate at home \citep{Shiftan1998}, in which household members depart together, engage in shared activities, and return together \citep{HoMulley2015}.
For brevity, fully-joint home-based tours are referred to simply as ``joint tours'' hereafter.
Typical examples include family shopping trips, diner trips as a couple, or trips to the park with young children.
These tours are governed by fundamentally different behavioural logics than solo travel: walking speeds may be constrained by the presence of young children, and public transit fares are charged per traveller, whereas fuel costs are shared.
Nonetheless, this difference is generally ignored in transport models and simulations, where joint and solo tours are handled with the same mode choice framework.
This can introduce systematic bias in preference parameter estimates \citep{HoMulley2015,HoMulley2015a}.

\todoi{aa: This fact that it introduces systematic bias is something just intuitive? Or is it also what you observe in your results? This is not clear from your phrasing. The use of word ``can'' is a bit too vague. I could say ``Sicily can explode due to the Etna volcano within 1000 years.'' However, in this case, this ``can'' denotes something theoretically possible, but whose plausibility is difficult to evaluate. If I say ``Le ressemblement national can win the election'': in this case ``can'' represents something that is likely to happen. In other word ``can'' denote too of a large spectrum of likelihood, and I think should be avoided here or at least clarified, since this ``bias'' is the motivation of the work.\\
lj: I justify my claim by citing two papers which investigate just that. Is it better?}
\todoi{aa: A bit better, but I don't think those cited papers talk explicitly about such a bias. If yes, please cite the precise subsection or page number in which they claim so. You can use \cite[\S X]{HoMulley2015}}

Several studies have examined mode choice in joint travel settings.
\cite{DissanayakeMorikawa2010} propose a combined revealed and stated preference Nested Logit model to jointly estimate vehicle ownership, mode choice, and trip-sharing decisions among couples in Bangkok.
\cite{DeLoachTiemann2012} apply a Multinomial Logit model to estimate mode shares across solo driver, fampool, formal carpool, and other modes in the United States, with a focus on fuel price sensitivity.
\cite{HoMulley2015} use a Nested Logit model to jointly estimate intra-household travel arrangements and mode choice in Australia.
\cite{PicardEtAl2018} model the joint mode choice of both spouses in French couples, explicitly accounting for the possibility of travelling together by car.

While these studies provide valuable behavioural insights, they stop short of producing outputs that can be directly integrated into transport simulations.
This paper addresses that gap by proposing a three-step pipeline --- joint tour classification, mode choice estimation, and driver/passenger assignment --- designed to produce simulation-ready trip assignments from standard survey and synthetic population data.
The resulting framework enables reliable evaluation of policies whose impacts differ between joint and solo travel, such as high-occupancy vehicle (HOV) lanes or family public transit fare discounts.


\section{Methodology}

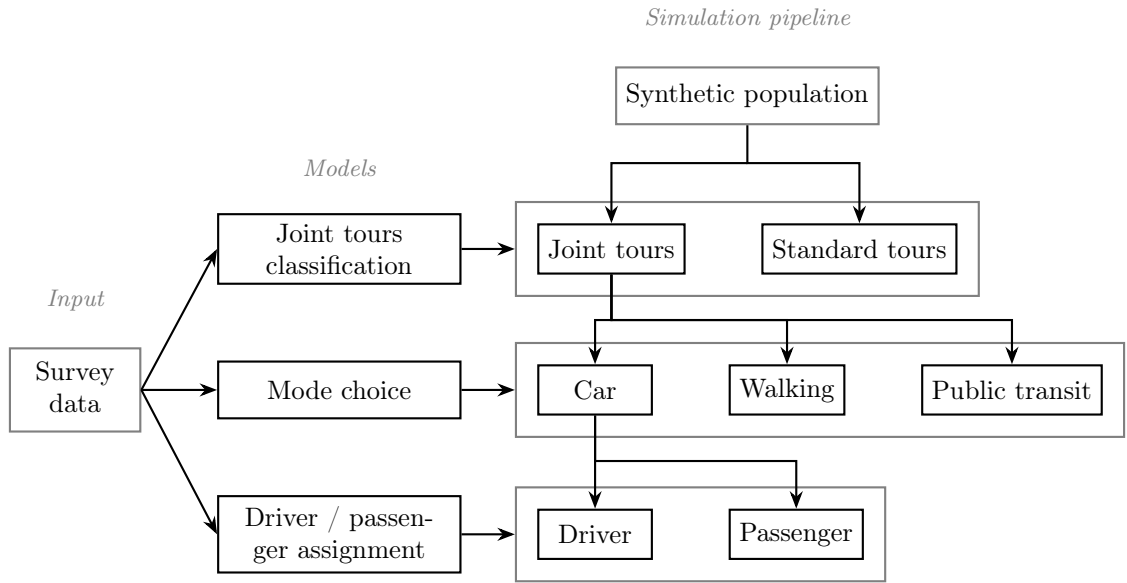
\begin{figure}[!h]
    \centering
    \begin{tikzpicture}[node distance=0.5cm]

        \node[mainbox, draw=gray, minimum width=0cm] (survey) {%
            \begin{tabular}{c}Survey\\data\end{tabular}};

        \node[mainbox, right=1cm of survey, text width=2.7cm]  (mlogit)    {Mode choice};
        \node[mainbox, above=1cm of mlogit, text width=2.7cm] (mlshared)  {Joint tours classification};
        \node[mainbox, below=1cm of mlogit, text width=2.7cm] (mldriver)  {Driver / passenger assignment};

        \node[nestedbox, right=1cm of mlshared]    (sharedtours)  {Joint tours};
        \node[nestedbox, right=1cm of sharedtours] (stdtours)     {Standard tours};
        \begin{scope}[on background layer]
            \node[groupbox, fit=(sharedtours)(stdtours),
                label={[label]above:Tour type}] (row2box) {};
        \end{scope}

        \node[mainbox, draw=gray, above=1cm of row2box] (synpop) {Synthetic population};

        \node[nestedbox, right=1cm of mlogit]     (sharedcar)  {Car};
        \node[nestedbox, right=1cm of sharedcar]  (sharedwalk) {Walking};
        \node[nestedbox, right=1cm of sharedwalk] (sharedpt)   {Public transit};
        \begin{scope}[on background layer]
            \node[groupbox, fit=(sharedcar)(sharedwalk)(sharedpt),
                label={[label]above:Mode split}] (row3box) {};
        \end{scope}

        \node[nestedbox, right=1cm of mldriver]   (shareddrv)  {Driver};
        \node[nestedbox, right=1cm of shareddrv] (sharedpax)  {Passenger};
        \begin{scope}[on background layer]
            \node[groupbox, fit=(shareddrv)(sharedpax),
                label={[label]above:Role split}] (row4box) {};
        \end{scope}


        \draw[arr] (survey.east) -- (mlshared.west);
        \draw[arr] (survey.east) -- (mlogit.west);
        \draw[arr] (survey.east) -- (mldriver.west);

        \draw[arr] (mlshared.east)  -- (row2box.west);
        \draw[arr] (mlogit.east)    -- (row3box.west);
        \draw[arr] (mldriver.east)  -- (row4box.west);

        \draw[arr] (synpop.south) ++(0,-0.5) -| (sharedtours.north);
        \draw[arr] (synpop.south) -- ++(0,-0.5) -| (stdtours.north);

        \draw[arr] (sharedtours.south) -- ++(0,-0.6) -| (sharedcar.north);
        \draw[arr] (sharedtours.south) -- ++(0,-0.6) -| (sharedwalk.north);
        \draw[arr] (sharedtours.south) -- ++(0,-0.6) -| (sharedpt.north);

        \draw[arr] (sharedcar.south) -- ++(0,-0.6) -| (shareddrv.north);
        \draw[arr] (sharedcar.south) -- ++(0,-0.6) -| (sharedpax.north);



        \node[label, above=0.35cm of survey]   {Input};
        \node[label, above=0.35cm of mlshared] {Models};
        \node[label, above=0.35cm of synpop]   {Simulation pipeline};

    \end{tikzpicture}
    \caption{Methodology to integrate joint tours in transport simulations}
    \label{fig:diagram}
\end{figure}

The proposed methodology to integrate joint home-based tours in a transport simulation is illustrated in Figure~\ref{fig:diagram}.
It takes two inputs: a \emph{household travel survey}, reporting travel diaries alongside household, person, and trip characteristics; and a \emph{synthetic population}, providing a random reproduction of the studied population with the trips to be simulated.
Three statistical models are estimated on the survey data and then applied sequentially to the synthetic population to produce trip assignments ready to be simulated.

We assume that household daily activity patterns are fixed, that is, the methodology addresses mode assignment for pre-specified activity chains rather than the generation of those chains.
Activity pattern generation is a distinct problem, addressed elsewhere in the literature by, for example, \cite{GliebeKoppelman2002}, \cite{MeisterEtAl2005}, or \cite{RezvanyEtAl2023}.

\subsection{Joint tour classification.}

The first model uses a Machine Learning (ML) classifier to determine whether each home-based tour is performed jointly with other household members or alone.
Joint tour status is not modelled as the outcome of an individual utility-maximising decision but rather as a latent tour characteristic to be predicted from observable predictors, making a machine learning classifier the more appropriate tool.

When applied to the synthetic population, the classifier produces for each home-based tour a predicted probability of being joint.
A Bernoulli draw is then performed for each tour using this predicted probability, determining whether the tour is classified as joint or not.

\subsection{Mode choice for joint tours.}

The second model uses a Multinomial Logit (MNL) model to estimate mode choice among car, walking, and public transit for joint tours.
Mode choice is modelled at the tour level, reflecting the empirical observation that travellers rarely mix modes within a tour (see below).
The car alternative encompasses both driver and passenger trips, since the mode decision is assumed to be made collectively among tour participants, with the driver/passenger assignment addressed subsequently by the third model.
Crucially, estimating mode choice separately for joint and solo tours avoids the parameter bias introduced by pooled models, which cannot account for the cost-sharing structure of car travel or the per-person fare structure of public transit.

\subsection{Driver/passenger assignment.}

The third model uses an ML classifier to assign each person in car-mode joint tour to either a driver or passenger role.
As with joint tour classification, an ML classifier is more appropriate since the goal is to predict the role of each traveller rather than to represent the underlying decision process.
In addition, simple natural rules can be used to assign automatically persons to a role, such as assigning as passenger all persons with no driving licence.
This step is crucial for transport simulations: only drivers contribute to road congestion, so accurate role assignment is necessary to properly estimate traffic volumes.

\subsection{Integration into agent-based simulation.}

The three models are estimated separately and then applied sequentially to the synthetic population within an agent-based transport simulator.
Tours are first partitioned into joint and non-joint groups with an ML classifier.
Then, joint and non-joint tours are handled by distinct mode choice models: joint tours use the MNL model presented above, while non-joint tours use the simulator's standard mode choice model.
Finally, joint car tours are assigned a driver or passenger role by an ML classifier.

This pipeline enables the simulator to differentiate policy impacts across joint and non-joint travel.
For instance, HOV lanes reduce car travel times specifically for joint tours, potentially inducing a modal shift away from public transit for those trips.
Conversely, family public transit fare discounts could shift joint tours from car to transit, an effect that cannot be captured in conventional simulations where co-traveller information is absent.
Properly modelling these mode shifts in transport simulations is crucial to evaluate urban policies.

One important point should be noted: tour classification is performed at the individual level, as in \cite{HoMulley2015}, meaning the model does not specify which household members travel together.
This is a practical necessity, as activity chains in the synthetic population are generated independently across household members.
Nevertheless, the model correctly replicates aggregate joint tour shares and mode distributions, which are the quantities of primary interest for policy evaluation.


\section{Results and discussion}
\todoi{aa: I think here you should discuss the types of models you use. The reviewer may say the usual boring stuff: ``the parameters of the models are not declared? what is the architecture of the neural network? How many layers? Is it long to train? How many trees in the random forest?''\\
Putting all these details in the paper would just make it look uglier. A solution could be that you put the code in github and you put a readme where you explain all these parameters. In the paper, you can say ``To ensure reproducibility, we provide a detailed and commented discussion of the parameters used in [github link], along with the code.''}

\subsection{Case study.}

The methodology is applied to the Paris region, home to approximately 12 million inhabitants.
The household travel survey used for model estimation is the \emph{Enquête Globale Transport} (EGT), conducted in 2010 \citep{EGT_2010}.
The EGT records household composition, individual sociodemographic characteristics, and all trips made by respondents on the day prior to the interview.

The data comprises \num{12452} fully responding households, representing \num{25686} individuals and \num{97191} trips.
Of the \num{36791} home-based tours identified, \num{4066} (\SI{11.1}{\%}) are classified as joint with another household member, defined as two or more persons reporting identical activity timings, locations, and purposes.

Table~\ref{tab:tour_mode_shares} presents mode shares separately for joint and non-joint home-based tours.
Only \SI{3}{\%} of home-based tours involve a mode change across trips within the tour, confirming that tour-level mode choice modelling is appropriate.
Mode distributions differ substantially between the two groups: car passenger tours are far more prevalent among joint tours than solo tours (\SI{33}{\%} vs. \SI{6}{\%}), while public transit and bicycle tours are considerably less common (\SI{9}{\%} vs. \SI{26}{\%} and $<$\SI{1}{\%} vs. \SI{3}{\%}, respectively).
Walking shares are similar across both groups.
These patterns confirm that joint tours warrant a dedicated mode choice model.
We restrict the study of the joint tour mode choice model to car, walking, and public transit, since together they account for \SI{97}{\%} of joint tours.

\begin{table}[ht!]
    \centering
    \caption{Mode shares for joint and non-joint home-based tours}
    \label{tab:tour_mode_shares}
    \begin{tabular}{lcc}
        Mode & Joint tours & Non-joint tours \\
        \hline
        Walking & \SI{32}{\%} & \SI{34}{\%} \\
        Car driver & \SI{23}{\%} & \SI{26}{\%} \\
        Car passenger & \SI{33}{\%} & \SI{6}{\%} \\
        Public transit & \SI{9}{\%} & \SI{26}{\%} \\
        Bicycle & $<$\SI{1}{\%} & \SI{3}{\%} \\
        Motorcycle & $<$\SI{1}{\%} & \SI{1}{\%} \\
        Other & $<$\SI{1}{\%} & $<$\SI{1}{\%} \\
        \emph{Mode combinations} & \SI{3}{\%} & \SI{3}{\%} \\
    \end{tabular}
\end{table}

The predictor variables used for the three models are reported in Table~\ref{tab:predictors}.
The synthetic population is generated following the methodology of \citet{HorlBalac2021}.

\begin{table}[!h]
    \centering
    \caption{Predictors used in the three models}
    \label{tab:predictors}
    \begin{tabular}{lccc}
        & Joint tour & Mode & Driv./pass. \\
        Variable & classif. & choice & assignment \\
        \hline
        \multicolumn{4}{l}{\emph{Tour level}} \\
        Number of activities & $\checkmark$ & & \\
        Activity duration & $\checkmark$ & & \\
        Trip distance & $\checkmark$ & & \\
        Activity start time & $\checkmark$ & & \\
        Activity end time & $\checkmark$ & & \\
        Leisure activity (0/1) & $\checkmark$ & $\checkmark$ & \\
        Shopping activity (0/1) & $\checkmark$ & $\checkmark$ & \\
        Work activity (0/1) & $\checkmark$ & & \\
        Education activity (0/1) & $\checkmark$ & & \\
        Other activity (0/1) & $\checkmark$ & & \\
        Travel time by mode & & $\checkmark$ & \\
                             & & & \\
        \multicolumn{4}{l}{\emph{Person level}} \\
        Age & $\checkmark$ & & $\checkmark$ \\
        Age gap to oldest & & & $\checkmark$ \\
        Gender & $\checkmark$ & & $\checkmark$ \\
        Driving licence (0/1) & $\checkmark$ & & $\checkmark$ \\
        Professional occupation & \multirow{2}*{$\checkmark$} & & \multirow{2}*{$\checkmark$} \\
        \footnotesize(4 groups) & & & \\
                                & & & \\
        \multicolumn{4}{l}{\emph{Household level}} \\
        Household structure & & & \multirow{2}*{$\checkmark$} \\
        \footnotesize(5 groups) & & & \\
        Nb.\ persons & $\checkmark$ & & \\
        Nb.\ minors & $\checkmark$ & $\checkmark$ & $\checkmark$ \\
        Nb.\ adult women & & & $\checkmark$ \\
        Nb.\ adult men & & & $\checkmark$ \\
        Nb.\ cars & $\checkmark$ & $\checkmark$ & $\checkmark$ \\
        Nb.\ driving licences & $\checkmark$ & & $\checkmark$ \\
    \end{tabular}
\end{table}


\subsection{Joint tour classification}
\label{sec:joint-tour-classification}

The first model aims to classify home-based tours as being joint or solo, by learning from the EGT data.
A Random Forest classifier is selected, outperforming seven alternative classifiers in cross-validation, with hyperparameters tuned through the same procedure.
The Random Forest naturally handles the mix of binary, categorical, and continuous features in the dataset.
Note that for \num{7110} tours where the person lives alone, the tour is automatically classified as solo, and only the remaining \num{29681} tours are passed to the classifier.

\begin{table}[ht!]
    \centering
    \caption{Performance metrics for the Random Forest model, compared to a stratified random baseline (mean $\pm$ std over 5 folds).}
    \label{tab:rf_metrics}
    \begin{tabular}{lcc}
        \textbf{Metric} & \textbf{Random Forest} & \textbf{Baseline} \\
        \hline
        Brier score (smaller is better) & $0.071\pm 0.001$ & $0.234\pm0.004$ \\
        Average Precision (larger is better) & $0.679\pm 0.011$ & $0.138 \pm 0.002$ \\
        F1 score (larger is better) & $0.547\pm 0.019$ & $0.133 \pm 0.017$ \\
    \end{tabular}
\end{table}

\begin{figure}[ht!]
    \centering
    \includegraphics{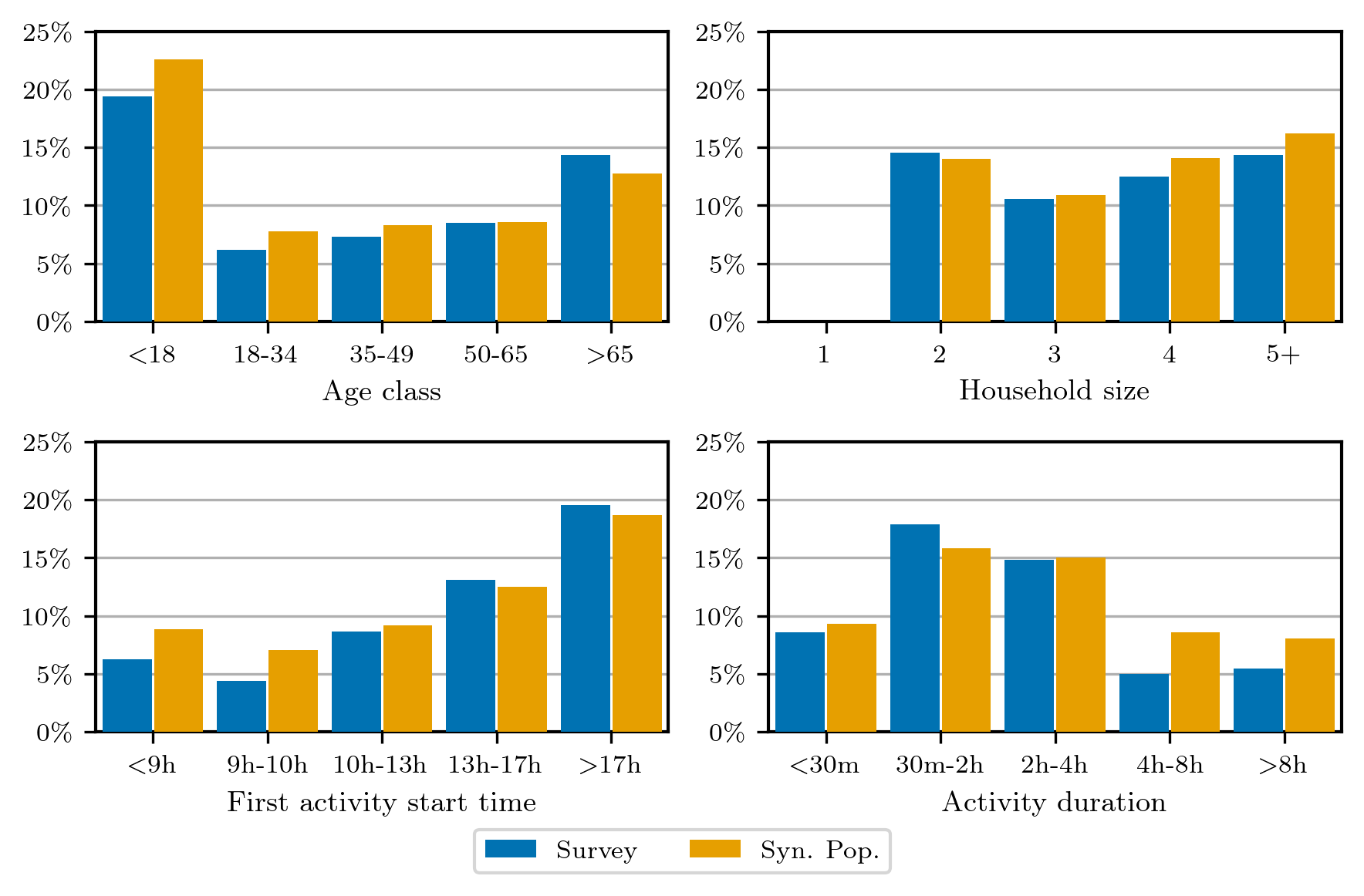}
    \caption{Ratio of joint tours in the travel survey and the synthetic population over four variables}
    \label{fig:synpop_shared_comparison}
\end{figure}

After training the classifier on the survey, we apply it to the tours of the synthetic population.
Figure~\ref{fig:synpop_shared_comparison} shows that the distribution of joint tours in the synthetic population replicates well the one observed in the survey  across four key variables: age, household size, first activity start time, and activity duration.
The share of total travelled distance attributable to joint tours is also well reproduced: \SI{7.5}{\%} in the survey versus \SI{9.0}{\%} in the synthetic population.

The strongest predictors of joint touring, from permutation importance analysis, are age, total distance between activities, start time of the first activity, total activity duration, and household size.
These results are broadly consistent with behavioural expectations: larger households with young children are more likely to take part in joint activities, later trips may reflect leisure purposes that are more typically shared, and activity duration may proxy for the type of activity undertaken.

To assess temporal stability, the model trained on EGT 2010 data was applied to EGT 2020 home-based tours as test set, yielding a still low Brier score of 0.101 suggesting that the determinants of joint touring are relatively stable over time.



\subsection{Mode choice for joint tours.}

For the \num{3920} joint tours in the EGT data whose mode is walking, car, or public transit, we estimate a Multinomial Logit model of mode choice using the Biogeme Python package \citep{Bierlaire2023}.
Travel time data are derived from three sources: walking travel times are computed from shortest-path distances on the OpenStreetMap network assuming a speed of \SI{3.5}{\kilo\meter/\hour}; public transit travel times are derived from GTFS data; and car travel times are taken from a METROPOLIS2 simulation \citep{Javaudin2024}, accounting for congestion at the trips' departure times.
Walking is used as the reference alternative.
The estimation results are presented in Table~\ref{tab:mnl_mode_choice}.

\begin{table}[!h]
    \centering
    \caption{Multinomial Logit estimation results for mode choice of joint tours}
    \label{tab:mnl_mode_choice}
    \begin{tabular}{lccc}
        & \multicolumn{3}{c}{Alternative} \\
        \hline
        Variable & Car & Public transit & Walking \\
        \hline
        Alternative-specific constant & $-3.78^{*}$ (17.3) & $-3.62^{*}$ (14.2) & -- \\
        Travel time (hour) & $-2.64^{*}$ (8.09) & $-1.52^{*}$ (6.91) & $-4.50^{*}$ (18.2) \\
        Leisure activity (0/1) & $-0.847^{*}$ (5.89) & $-0.737^{*}$ (3.88) & -- \\
        Shopping activity (0/1) & $-0.033$ (0.229) & $-0.607^{*}$ (3.19) & -- \\
        Number of cars & $1.27^{*}$ (15.8) & $-0.485^{*}$ (3.65) & -- \\
        Number of minors & $-0.077$ (1.60) & 0.101 (1.66) & -- \\
        \hline
        \multicolumn{4}{l}{\textit{Model fit}} \\
        Number of observations & \multicolumn{3}{c}{3920} \\
        Null log-likelihood & \multicolumn{3}{c}{$-4270.1$} \\
        Final log-likelihood & \multicolumn{3}{c}{$-1772.5$} \\
        Adjusted $\rho^2$ & \multicolumn{3}{c}{0.762} \\
    \end{tabular}
    \small
    \vspace{2ex}

    Each cell reports the estimated coefficient with the $t$-statistic in parentheses.\\
    $^{*}p<0.05$
\end{table}

The alternative-specific constants for both car and public transit are negative relative to walking.
For car, this likely reflects fixed costs associated with parking and vehicle access at the destination.
For public transit, it may partly capture the fare cost, which is not included as a separate variable in the utility function.

Travel time has a negative effect on utility for all three modes, as expected.
The disutility of travel time is largest for walking (\num{4.50} per hour) and smallest for public transit (\num{1.52} per hour), suggesting that travellers are more tolerant of time spent in a vehicle than time spent on foot, a finding consistent with the broader mode choice literature.

Performing at least one leisure activity during the tour reduces the utility of car and public transit relative to walking.
This is consistent with some leisure tours that are pedestrian in nature, such as a walk to a park, where walking is itself part of the activity.
Performing at least one shopping activity reduces the utility of public transit relative to walking, plausibly reflecting the inconvenience of carrying purchased goods on board.
The number of cars owned by the household has a positive effect on car utility and a negative effect on public transit utility, consistent with the role of vehicle availability as a key determinant of mode choice.
The number of minors in the household does not reach statistical significance for either alternative, suggesting that, conditional on the other covariates, household composition does not substantially affect mode choice for joint tours.


\subsection{Driver/passenger assignment.}

Among the \num{2235} joint car tours in the EGT data, \SI{40}{\%} are performed as driver and \SI{60}{\%} as passenger.
For \num{914} of them, the role can automatically be assigned: persons with no driving licence are assigned as passengers and persons who hold the only driving licence in the households are assigned as drivers.

For the remaining \num{1321} joint car tours, a ML model is used to classify the persons as either driver or passenger, based on the predictors listed on Table~\ref{tab:predictors}.
Eight classifiers were evaluated using the same cross-validation protocol as for joint tour classification.
The Logistic Regression with L2 penalty achieves the best performance in terms of Brier score and Average Precision (Table~\ref{tab:mlp_metrics}), and is therefore selected for driver/passenger assignment.

\begin{table}[ht!]
    \centering
    \caption{Performance metrics for the Logistic Regression classifier, compared to a stratified random baseline (mean $\pm$ std over 5 folds).}
    \label{tab:mlp_metrics}
    \begin{tabular}{lcc}
        Metric & Logistic Regression & Baseline \\
        \hline
        Brier score (smaller is better) & $0.145\pm 0.017$ & $0.482\pm0.047$ \\
        Average Precision (larger is better) & $0.833\pm 0.033$ & $0.553 \pm 0.032$ \\
        F1 score (larger is better) & $0.825\pm 0.029$ & $0.559 \pm 0.028$ \\
    \end{tabular}
\end{table}

\begin{figure}[!h]
    \centering
    \includegraphics{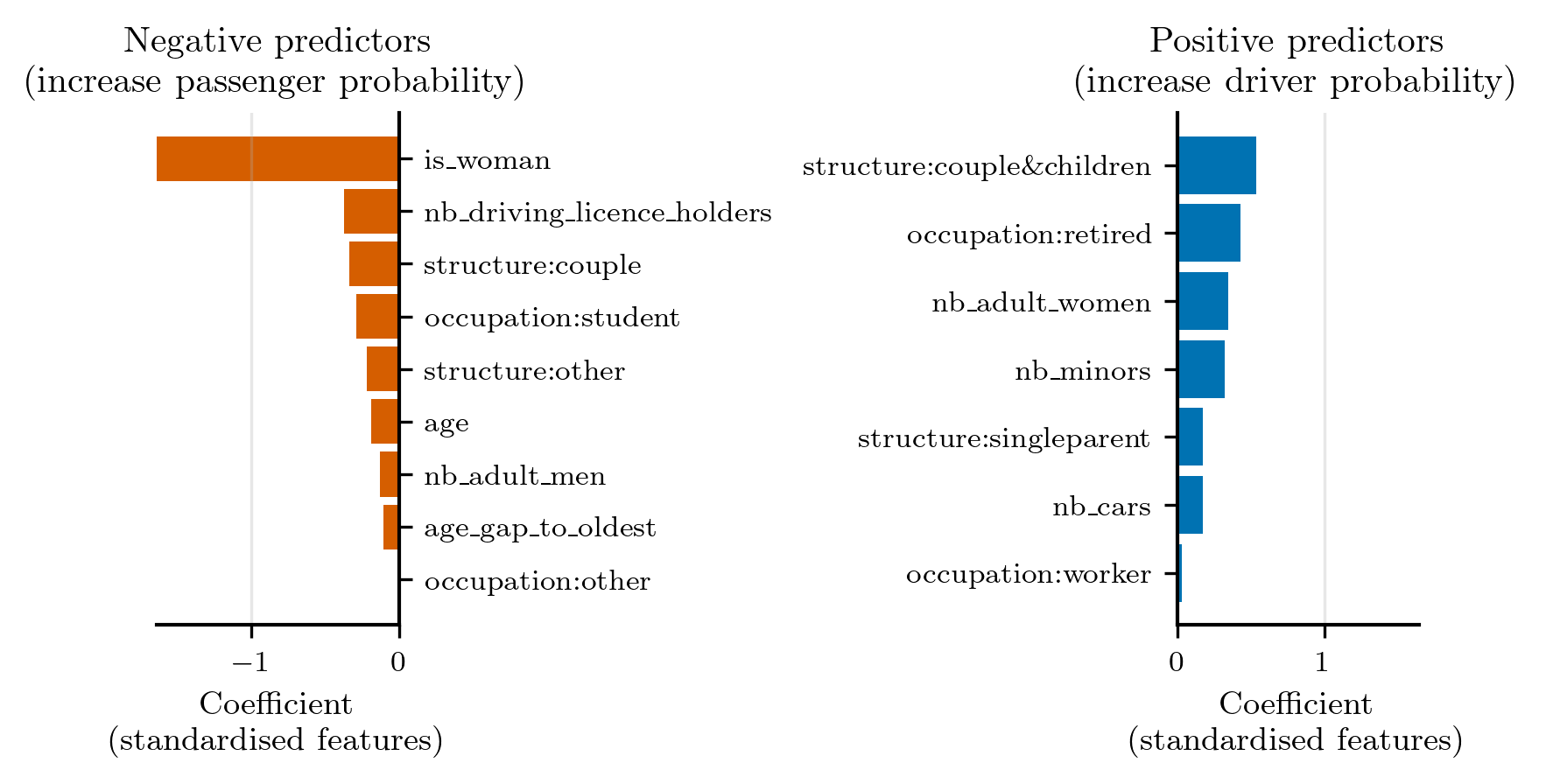}
    \caption{Predictors coefficients of the Logistic Regression}
    \label{fig:lr_coefs}
\end{figure}

Figure~\ref{fig:lr_coefs} presents the standardised coefficients of the Logistic Regression model.
Among the strongest predictors, being a woman substantially increases the probability of being a passenger, consistent with empirical evidence that men are more likely to drive in couple travel.
Additionally, belonging to a couple-with-children household increases the probability of being a driver, suggesting that in such households one parent is often traveling with the children but without the other spouse, and must therefore take the driver role.

\section{Conclusions}

This paper proposes a methodology to integrate intra-household fully-joint home-based tours in agent-based transport models.
Three statistical models are estimated sequentially on household travel survey data and applied to a synthetic population: a Random Forest classifier to identify joint tours, a Multinomial Logit model for mode choice specific to joint tours, and a Penalized Logistic Regression for driver/passenger assignment.

Properly representing joint tours in transport simulation has direct policy relevance.
Policies whose impacts differ systematically between joint and solo travel, such as HOV lanes or family public transit fare discounts, cannot be reliably evaluated in conventional simulations where joint-trip information is absent.
The proposed methodology provides the modelling infrastructure needed to assess such policies more accurately.

Several limitations point to directions for future work.
First, tour classification is performed at the individual level, meaning the model replicates aggregate joint tour shares without specifying which household members travel together.
Extending the methodology to explicitly match household members would enable a richer representation of intra-household coordination.
Second, the methodology assumes fixed activity patterns, and integrating joint tour generation with activity scheduling models remains an open challenge.

\section*{Acknowledgements}

This work has benefited from invaluable advise and feedback from Martin Briand, Pierre Freval, and Tatiana Seregina.
Any errors or omissions that may remain are the sole responsibility of the author.

This work has been carried out at the Energy4Climate Interdisciplinary Center (E4C) of IP Paris, which is in part supported by 3rd Programme d’Investissements d’Avenir [ANR-18-EUR-0006-02].
This work was supported by funding from the French National Research Agency (ANR) under the France 2030 program (grant reference: “ANR-24-PEMO-0003”).

\bibliography{mylibrary.bib}

\end{document}